\documentclass[multphys]{svmult}

\usepackage{amsmath}
\usepackage{graphicx}
\usepackage{calrsfs}

\DeclareGraphicsExtensions{.eps}

\begin{document}

\title*{Spin- and Energy Relaxation of Hot Electrons at GaAs Surfaces}
\author{T. Ohms \and K. Hiebbner \and H. C. Schneider
\and M. Aeschlimann}
\institute{Department of Physics\\ Kaiserslautern University of
Technology\\ 67663 Kaiserslautern \\Germany}

\begin{abstract}
The mechanisms for spin relaxation in semiconductors are reviewed, and
the mechanism prevalent in p-doped semiconductors, namely spin
relaxation due to the electron-hole exchange interaction, or
Bir-Aronov-Pikus mechanism, is presented in some depth. It is shown
that the solution of Boltzmann type kinetic equations allows one to
obtain quantitative results for spin relaxation in semiconductors that
go beyond the original Bir-Aronov-Pikus relaxation-rate
approximation. Experimental results using surface sensitive two-photon
photoemission techniques show that the relaxation time of the elctron
spin polarization in p-doped GaAs at a semiconductor/metal surface is
several times longer than the corresponding bulk spin relaxation
times. A theoretical explanation of these results in terms of the
reduced density of holes in the band-bending region at the surface is
presented.

\end{abstract}

\maketitle

\section{Introduction}

Semiconductor technology has relied on the manipulation of electronic
charges since the invention of the transistor. The spin degree of
freedom of the carriers has been mostly ignored in traditional
electronics. In the last few years there has been a push towards the
control of the spin dynamics of charged carriers
\emph{independently} of their charge. The exploitation of the spin
degree of freedom in
electronic~\cite{prinz:science:magnetoelectronic-devices} and
opto-electronic devices~\cite{sham:science:device} and the basic
physics associated with the control of spins have been dubbed
``spintronics.''

Outside of semiconductor electronics, the control of spins has gained
tremendous importance in basic and applied physics in
connection with magnetic recording
techniques.~\cite{akinaga-ohno:review} Metallic materials, which are
routinely used for these purposes, have some disadvantages compared to
semiconductors.  For instance, their carrier densities are not freely
controllable and they do not possess bandgaps, which makes their use
in opto-electronic devices difficult. It would therefore be
advantageous to combine some features of magnetic materials with the
versatility of semiconductors. However, the decay of the spin
polarization in semiconductors limits all information processing and
storage capabilities in semiconductors, which is a big difference from
conventional electronics that is based on the conserved electronic
charge.  Spin relaxation and methods to control spin relaxation thus
constitute central problems of spintronics. A promising result on the
physical limit of spin manipulation by relaxation phenomena is the
measurement of a spin lifetime of about 1 nanosecond in n-doped GaAs
at low temperatures~\cite{kikkawa:prl98}, but the reason for this
remarkable result remains an area of active research. 

One important requirement for spin-electronic devices, such as the
spin transistor, is the efficient injection of spin-polarized carriers
into semiconductors. The straightforward implementation by using
ferromagnetic metal contacts has proved to be difficult mainly due to
the different conductivities of ferromagnetic metals and
semiconductors~\cite{schmidt:prb00}, but tunnel injection into
non-magnetic semiconductors was realized by using incoherent
electrical tunneling into the target semiconductor from ferromagnetic
metals~\cite{jonker:procieee03}: In tunneling contacts the injected
current is proportional to the density of states of the respective
material. Different densities of states for spin-up and spin-down
electrons in the ferromagnet therefore give rise to a spin-polarized
tunneling current leading to injected spin polarizations of up to
30\%. An alternative to the spin injection via interfaces between
ferromagnetic metals and semiconductors is the use of magnetic
semiconductors instead of ferromagnets. For instance, doping II-VI and
III-V semiconductors with Mn can lead to paramagnetic, e.g., BeMnZnSe,
and even ferromagnetic compounds such as
GaMnAs~\cite{awschalom-book}. Aside from the intrinsic importance of
these structures, they can be used for spin injection into
non-magnetic
semiconductors~\cite{akinaga-ohno:review,awschalom:prb02:zener-tunneling}.
However, these techniques suffer from characteristic drawbacks. For
instance, paramagnetic semiconductors based on II-VI compounds,
require high magnetic fields. The ferromagnetic GaMnAs, on the other
hand, has a relatively low Curie temperature of 150\,K and works with
spin aligned holes, for which momentum-space scattering in bulk (and
quantum-well) semiconductors severely limits the lifetimes. Since
these problems with electrical spin injection persist, optical
techniques are important not only for probing the spin alignment of
the carrier system, but also for the creation of a well-controlled
spin alignment at arbitrary temperatures. The experimental techniques
described in the following rely on the creation of spin-polarized
electrons by optical fields and the emission of carriers from the
material using laser pulses, and have the advantage that the
spin-polarization can be obtained directly from the carriers, which
are freed from the material.

In this contribution, we briefly review some of the mechanisms that
lead to the relaxation of spin polarization in semiconductors. We then
focus on p-doped GaAs and its prevalent spin-relaxation mechanism, the
electron-hole exchange interaction, to show that the relaxation of the
spin polarization is determined by the full dynamics of the electronic
distribution functions and cannot be described using a simple
relaxation rate. Instead, we present theoretical results for the
spin-polarization dynamics in p-doped semiconductors using a Boltzmann
equation approach. In a second part of this contribution we show how
the experimental technique of two-photon photoemission (2PPE) can be
used to extract information about spin dynamics at semiconductor
surfaces and how it can be applied to semiconductor-metal interfaces
like Schottky contacts. Experimental results are presented and
theoretically explained.

\section{Review of Spin-Flip Processes in GaAs}

\label{ma:sec-theory}

In this section we first give an overview of the electron and hole
states accessible by optical excitation of GaAs, and the possibility
to create spin-polarized electrons by optical excitation. We then
describe the origin of the three most important processes that lead to
spin relaxation. Since they are all connected to the bandstructure of
semiconductors, we first discuss some generalities on the
semiconductor bandstructure, and how it can be approximately
calculated close to center of the Brillouin zone.

\subsection{Optical Orientation of Photo-Excited Carriers}

\begin{figure}[t]
\begin{center}
\resizebox{0.8\textwidth}{!}{\includegraphics*{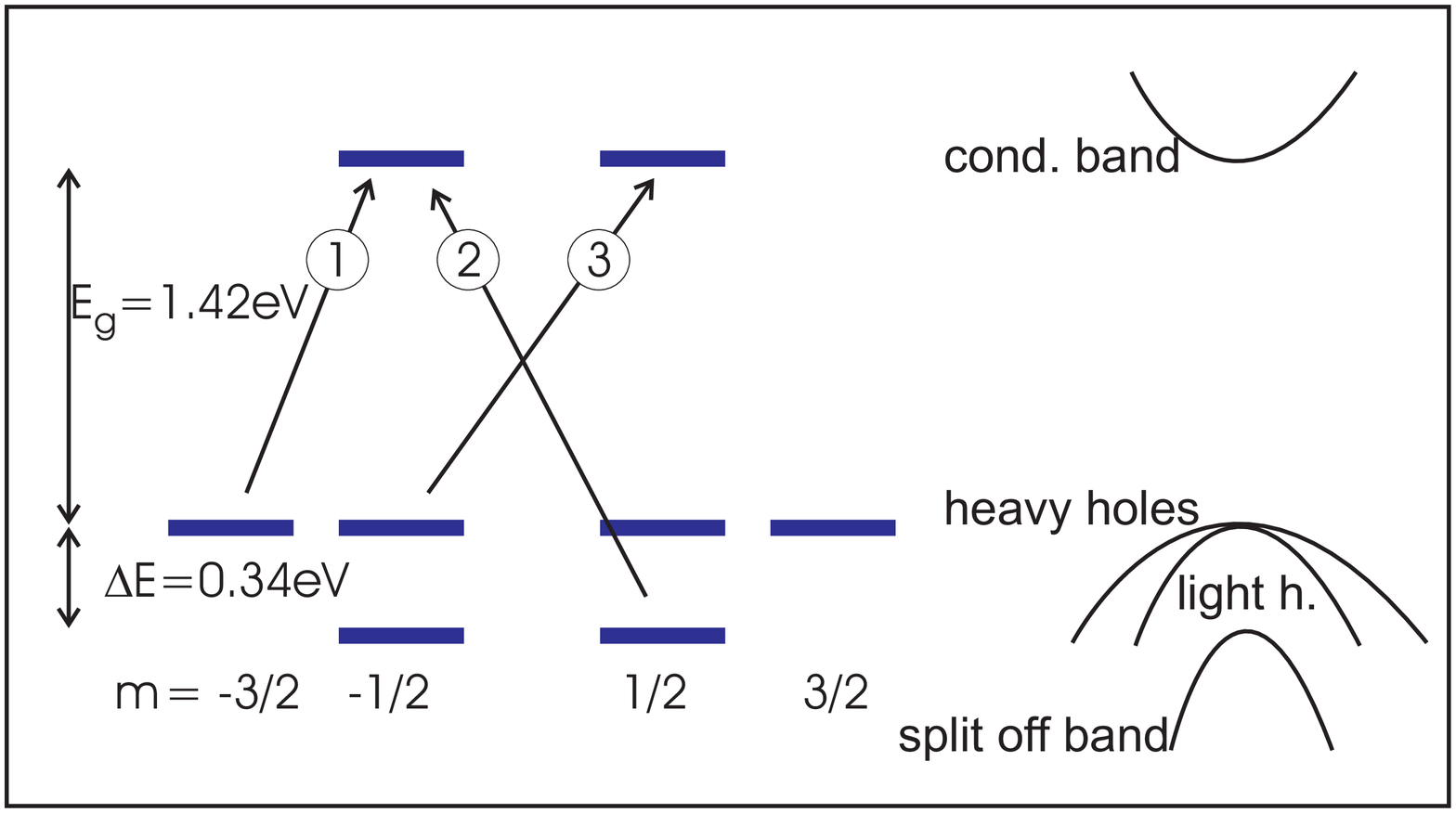}}
\caption{Transitions between electron and hole states induced by
circularly polarized light in GaAs at the zone center (left) and
schematic bandstructure (right). The angular momentum projection
quantum numbers $m$ shown in the figure apply to all bands above the
respective $m$s. The relative strengths of the transitions are
indicated. For excitation photon energies of less than 1.76\,eV, only
heavy and light-hole transitions can be driven.}
\label{ma:GaAs-transitions}
\end{center}
\end{figure}

Optical orientation refers to the creation of a non-equilibrium spin
alignment, or preferential spin orientation, by excitation with
polarized electromagnetic fields. As mentioned above, in
semiconductors this is the most important and versatile process to
create spin alignment, or as it is commonly called, spin
polarization. In the following, we investigate the microscopic
\emph{spin polarization}
\begin{equation}
P=\frac{n_{\uparrow}-n_{\downarrow}}{n_{\uparrow}+n_{\downarrow}}
\label{ma:pol}
\end{equation}
defined in terms of the microscopic, time and momentum (or kinetic
energy) dependent carrier densities in the spin-up and spin-down
electron bands. The dynamics of the microscopic spin polarization $P$
determines the relaxation of the macroscopic spin polarization, which
is often described by a phenomenological $T_1$ time. We therefore
refer to the \emph{decay of the electron spin-polarization simply as spin
relaxation}. The experimental and theoretical results on spin
relaxation presented in the following are obtained without external
magnetic fields, and should therefore not be confused with the
dephasing of \emph{coherent} spin dynamics under the influence of
magnetic fields~\cite{crooker:prb97,hallstein:prb97}, whose
macroscopic counterpart is often described by a $T_2$ time, see also
Sec.~\ref{ma:sub-pol-dynamics}. Spin-polarized carriers can be
detected in experiments and exploited in electro-optical devices due
to the polarization of the emitted light when spin-polarized electrons
and holes recombine and emit photons. The bandstructure of GaAs near
the fundamental band edge~\cite{yu-cardona} that will be the basis of
the following discussion consists of:
\begin{itemize}
\item electrons with total spin $S=\frac{1}{2}$ and spin projection quantum
number $s=+\frac{1}{2}\equiv\uparrow$ and $s=-\frac{1}{2}\equiv\downarrow$,
\item heavy holes with total angular momentum $J=\frac{3}{2}$ and projection
quantum number $j=\pm\frac{3}{2}$,
\item light holes with total angular momentum $J=\frac{3}{2}$ and projection
quantum number $j=\pm\frac{1}{2}$,
\item holes in the ``split-off'' band with
total angular momentum $J=\frac{1}{2}$ and projection quantum number
$j=\pm\frac{1}{2}$,
\end{itemize}
as shown schematically in Fig.~\ref{ma:GaAs-transitions}.
More specifically, one has for the spherically symmetric
conduction-band wavefunctions at $k=0$ in real space representation
\begin{equation}
\langle\vec{x},\sigma|S,s\rangle=u_{c}(r)\chi_{s}(\sigma)\ ,\qquad
\sigma=\uparrow,\downarrow \label{ma:uc}%
\end{equation}
where $r=|\vec{x}|$ is the modulus of $\vec{x}$.
The light and heavy hole valence band wavefunctions at $k=0$ are given by
\begin{align}
\langle\vec{x},\sigma|J=\frac{3}{2},j=  &  \frac{3}{2}\rangle=Y_{+1}^{1}%
(\hat{\vec{x}})\chi_{\uparrow}(\sigma)u_{v}(r)\label{ma:uv-1}\ ,\\
\langle\vec{x},\sigma|J=\frac{3}{2},j=  &  \frac{1}{2}\rangle=\frac{1}%
{\sqrt{3}}\left[  \sqrt{2}Y_{0}^{1}(\hat{\vec{x}})\chi_{\uparrow}(\sigma)
+Y_{+1}^{1}(\hat{\vec{x}})\chi_{\downarrow}(\sigma)\right]  u_{v}(r) \ ,
\label{ma:uv-2}\\
\langle\vec{x},\sigma|J=\frac{3}{2},j=  &  -\frac{1}{2}\rangle=\frac{1}%
{\sqrt{3}}\left[  \sqrt{2}Y_{0}^{1}(\hat{\vec{x}})\chi_{\downarrow}(\sigma
)+Y_{-1}^{1}(\hat{\vec{x}})\chi_{\uparrow}(\sigma)\right]  u_{v}(r)\ ,\label{ma:uv-3}\\
\langle\vec{x},\sigma|J=\frac{3}{2},j=  &  -\frac{3}{2}\rangle=Y_{-1}^{1}%
(\hat{\vec{x}})\chi_{\downarrow}(\sigma)u_{v}(r)\ , \label{ma:uv-4}%
\end{align}
and for the split-off hole wave functions one has
\begin{align}
\langle\vec{x},\sigma|J=\frac{1}{2},j=  &  \frac{1}{2}\rangle=
\frac{1}{\sqrt{3}}\left[  \sqrt{2}Y_{1}^{1}(\hat{\vec{x}})
   \chi_{\downarrow}(\sigma)-Y_{0}^{1}(\hat{\vec{x}})
   \chi_{\uparrow}(\sigma)\right]  u_{v}(r) \ ,
  \label{ma:so1}\\
\langle\vec{x},\sigma|J=\frac{1}{2},j=  &  -\frac{1}{2}\rangle =
\frac{1}{\sqrt{3}}
\left[  Y_{0}^{1}(\hat{\vec{x}})\chi_{\downarrow}(\sigma)
-\sqrt{2}Y_{-1}^{1}(\hat{\vec{x}})\chi_{\uparrow}(\sigma)\right]u_{v}(r)\ .
   \label{ma:so2}
\end{align}
Here, the $Y_{m}^{l}$ are the spherical harmonics that depend on the
direction~$\hat{\vec{x}}$ of the vector~$\vec{x}$, $\chi_{s}(\sigma
)=\delta_{\sigma,s}$ are spinors, and $u_{c}(r)$, $u_{v}(r)$ contain
the radial dependences, which can be obtained from a band-structure
calculation. From~(\ref{ma:uc}) and (\ref{ma:uv-1})--(\ref{ma:so2})
the vectorial dipole matrix elements $\vec{d} =e\vec{x}$ for
electron-hole transitions can be computed. For instance, choosing the
angular momentum quantization axis $\hat{z}$ perpendicular to the
crystal surface and assuming excitation with polarized light
propagating in $z$ direction, one has to evaluate the matrix elements
\begin{equation}
\langle S,s|\hat{\sigma}_{\pm}|J,j\rangle
\end{equation}
between electron and hole wavefunctions and the
polarization vector $\hat{\sigma}_{\pm}=(\hat{\vec{x}}\pm
\imag\hat{\vec{y}})/\sqrt{2}$. Because $\hat{\sigma}_{\pm}\propto
Y_{\pm1}^{1}$ this is relatively easy to do. For example, one obtains
for the ratio between electron to heavy hole and electron to
light-hole transitions
\begin{equation}
\left\vert \frac{\langle S=\frac{1}{2},\downarrow|\hat{\sigma}_{+}|J=\frac
{3}{2},j=-\frac{3}{2}\rangle}{\langle S=\frac{1}{2},\downarrow|\hat{\sigma
}_{+}|J=\frac{3}{2},j=-\frac{1}{2}\rangle}\right\vert ^{2}=3 \ .
\end{equation}
These relative magnitudes are indicated in Fig.~\ref{ma:GaAs-transitions}. It
is apparent that selective excitation of light and heavy hole transitions will
result in an electronic spin polarization of $P=0.5$. In the following, we
always assume that an electronic spin polarization is created in this way, and
that no split-off holes are populated in the optical excitation process. We
will therefore use the abbreviations
\begin{equation}
|j\rangle\equiv|J=\frac{3}{2},j\rangle\mathrm{, }|s\rangle\equiv|S
=\frac{1}{2},s\rangle. \label{ma:js-states}
\end{equation}

Once a spin-polarization is created by optical fields, several
mechanisms are known to destroy the spin-polarization. These
spin-relaxation mechanisms have been investigated in detail for bulk
and, more recently, for quantum-well semiconductors. In the following,
we will briefly review the Elliott-Yafet (EY), Dyakonov-Perel (DP),
and Bir-Aronov-Pikus (BAP) mechanisms, after some general properties
of bandstructure calculations near the fundamental band edge have been
discussed.

\subsection{Bandstructure Properties}

\label{ma:sub-bandstructure}

The bandstructure of III-V compounds is generally simple at the zone
center ($k=0$). For GaAs, the zone-center wavefunctions of the bands
at the fundamental banggap are given by
(\ref{ma:uc})--(\ref{ma:uv-4}). For finite $k$ the energy eigenstates
and dipole matrix elements for the bands participating in optical
transitions can be determined by perturbation theory starting from the
$k=0$ states. This is usually done by dividing the bands into those
energetically close to the transitions of interest and remote
bands. In the case of the interaction with optical fields, the
important bands are the heavy hole, light hole, and electron
states. Following~\cite{bir-pikus:book,ivchenko-pikus:book,winkler:book},
our starting point is the general form of single-particle carrier
states in a periodic potential
\begin{equation}
   \psi_{n,\vec{k}}(\vec{x},\sigma)=\frac{1}{\sqrt{L^{3}}}
   \E^{\imag\vec{k}\cdot\vec{x}}u_{n,\vec{k}}(\vec{x},\sigma)
\label{ma:Bloch}
\end{equation}
with the crystal volume $L^{3}$ and the lattice periodic Bloch
function $u_{n,\vec{k}}=|n,\vec{k}\rangle$. The indices $n$ run over all
bands in the semiconductor. A bandstructure calculation in general
attempts an approximate solution of
\begin{equation}
H_{\mathrm{carrier}}\psi_{n,\vec{k}}=\left[  \frac{1}{2m_{0}}p^{2}
+V_{\mathrm{lattice}}+H_{SO}\right]
\psi_{n,\vec{k}}=\epsilon_{n,\vec{k}}
\psi_{n,\vec{k}}
\label{ma:H-carr}
\end{equation}
where $H_{\mathrm{SO}}\propto\vec{L}\cdot\vec{S}$ is the spin-orbit
interaction, $\vec{p}=-i\hbar\nabla$ the linear momentum operator and
$V_{\mathrm{lattice}}$ the periodic potential of the crystal
lattice. 

\noindent Equations~(\ref{ma:Bloch}) and~(\ref{ma:H-carr}) lead directly to
\begin{equation}
   \left[  H_{\mathrm{carrier}}+H_{kp} +\frac{\hbar^2}{2m_0}k^2 \right]  
   |n,\vec{k}\rangle
   =\epsilon_{n,\vec{k}}|n,\vec{k}\rangle
   \label{ma:H-kp1}
\end{equation}
with
\begin{equation}
H_{kp}=\frac{\hbar}{m_{0}}\vec{k}\cdot\vec{p}\ .
\end{equation}
At $k=0$, the electron and hole states of interest are eigenstates of
the total angular momentum operator and given by~(\ref{ma:uc})
and~(\ref{ma:uv-1})--(\ref{ma:so2}). This set of states will be denoted by
$\mathcal{D}$ in the following. Together with the remote bands at
$k=0$, the states $\mathcal{D}$ satisfy
\begin{equation}
H_{\mathrm{carrier}}|n\rangle=\epsilon_{n}|n\rangle
\end{equation}
where $|n\rangle=|n,k=0\rangle$ and $\epsilon_n
=\epsilon_{n,\vec{k}=0}$. In the vicinity of $k=0$ one can now
determine the carrier states $|n,k\rangle$ by including the coupling
of the $\mathcal{D}$ states with the remote bands via~(\ref{ma:H-kp1})
as a 2nd-order perturbation term with interaction matrix element
\begin{equation}
\langle n'|H_{kp}(k)|n\rangle=\frac{\hbar}{m_{0}}\vec{k}\cdot\vec
{p}_{n'n} 
\end{equation}
where 
\begin{equation}
  \vec{p}_{n'n}=\langle n'| \vec{p}|n'\rangle \ .
  \label{ma:p-matrix-element}
\end{equation}
Expanding the states at $\vec k\neq0$ in terms of the states at $\vec k = 0$ 
\begin{equation}
|n,\vec{k}\rangle = \sum_m c_{n,m}(\vec{k})\, |n\rangle
\label{ma:basis-n}
\end{equation}
leads to the  Hamiltonian matrix
\begin{align}
  \mathcal{H}_{mn}(\vec{k})= &\left(\epsilon_n
  +\frac{\hbar^{2}}{2m_{0}}k^{2}\right)\delta_{m,n} \nonumber \\ 
 &\mbox{}+\frac{1}{2}\frac{\hbar^{2}}{m_{0}^{2}}
  \sum_{r\neq m,n}\sum_{\alpha,\beta
    =x,y,z}p_{n,r}^{\alpha}~p_{r,m}^{\beta}\left[  \frac{1}{\epsilon_{m}%
-\epsilon_{r}}+\frac{1}{\epsilon_{n}-\epsilon_{r}}\right]  \label{ma:luttinger}%
\end{align}
determining the single-particle energies $\epsilon_{n,\vec{k}}$ and
expansion coefficients in~(\ref{ma:basis-n}) via the eigenvalue
problem
\begin{equation}
  \sum_{m\in \mathcal{D}} \mathcal{H}_{n'm}(\vec{k})\, c_{m,n}(\vec{k}) 
  = \epsilon_{n,\vec{k}}\,c_{n'n}(\vec{k})\ .
  \label{ma:luttinger-matrix}
\end{equation} 
In~(\ref{ma:luttinger}), $p_{n'n}^{\alpha}$ is the Cartesian $\alpha$
component of $\vec{p}_{n'n}$ in~(\ref{ma:p-matrix-element}).  The
approach outlined above is known as $k\cdot p$ theory, but the
Hamiltonian~(\ref{ma:luttinger}) is usually replaced by an effective
Kohn-Luttinger Hamiltonian containing only a few parameters that can
be fitted to experimental results instead of the momentum matrix
elements with all the remote
bands~\cite{bir-pikus:book,yu-cardona,winkler:book}. In the present
case, the general band indices $m$, $n$ represent the hole quantum
numbers $j$ or the electron quantum numbers $s$. By
diagonalizing~(\ref{ma:luttinger-matrix}) for each $\vec{k}$, one
obtains the band-structure $\epsilon_{n}(k)$ and the coefficients
$c_{n'n}(\vec{k})$ determining the states $|n,k\rangle$
via~(\ref{ma:basis-n}). The advantages of this method to obtain the
bandstructure close to $k=0$, are that it avoids a full band-structure
calculation, and the effective Hamiltonian can be conveniently
parametrized in terms of a few parameters.

The procedure outlined in the previous paragraph yields an effective
Hamiltonian that describes the energy splitting of the spin-up and
spin-down conduction electron bands in certain crystal directions, and
therefore directly yields the Dyakonov-Perel process as will be
discussed below. However, it is important to note that this procedure
for obtaining effective Hamiltonians is not restricted to the original
Hamiltonian (\ref{ma:H-kp1}), but can also be generalized for
additional interactions important to the dynamics, such as the Coulomb
interaction, the interaction with phonons, or with
impurities. Depending on the form of the interaction this leads to
additional contributions to the effective Hamiltonian
$\mathcal{H}_{nn'}(\vec{k})$ of the general form
$\mathcal{H}^{\mathrm{ph}}_{nn'}(k,k')$ for phonon (or impurity)
interaction, which describes scattering with carrier states
$(n,k)\rightarrow(n',k')$. For the Coulomb interaction, the effective
Hamiltonian is an effective two-particle interaction of the form
$\mathcal{V}_{n_{1}n_{1}'n_{2}n_{2}'}(\vec{k}_{1},
\vec{k}_{1}',\vec{k}_{2},\vec{k}_{2}')$, which describes the
transition, or scattering process,
$(n_{1}',k_{1}')(n_{2}',k_{2}')\rightarrow
(n_{1},k_{1})(n_{2},k_{2})$.  As will be shown below, incorporating
the Coulomb interaction leads to the exchange interaction between
electrons and holes, i.e., the Bir-Aronov-Pikus process, and
carrier-phonon scattering leads to the Elliott-Yafet process.

\subsection{Elliot-Yafet Mechanism}

\label{ma:sub-EY}

The Elliot-Yafet mechanism is a spin-flip processes due to the
coupling between electrons and holes combined with phonon scattering
processes, which can be described by the following effective
Hamiltonian~\cite{pikus-titkov:opt-orient} including terms to 3rd
order in the subspace of the interesting electron bands
($s,s'=\uparrow$ or $\downarrow$)
\begin{equation}
  \mathcal{H}^{\mathrm{ph}}_{s's}(k',k)=
 \sum_{j}\frac{\mathcal{H}_{s'j}(\vec{k})\langle
 j|H_{e-p}(k',k)|j\rangle \mathcal{H}_{js}(\vec{k})}
 {(\epsilon_{j}-\epsilon_{s})^{2}}
     \label{ma:H-phon}
\end{equation}
where $H_{e-p}$ is the Fr\"{o}hlich Hamiltonian~\cite{kittel:qts} that
describes the long-range interaction with phonons. The Hamiltonian
matrix~(\ref{ma:H-phon}) includes the effect the coupling due to the
nonvanishing matrix elements~$\mathcal{H}_{sj}(\vec{k})$,
cf.~(\ref{ma:luttinger}), between electron and hole states. The phonon
scattering process changes the hole momentum $k\rightarrow k'$, and
therefore, in effect, also changes the mixture of electron and hole
states described by the overlap $\mathcal{H}_{sj}(\vec{k})$ since the
latter depends on the wave vector $k$. The spin-dependent electronic
lifetime\footnote[1]{With the factor of 2 included on the LHS
of~(\protect\ref{ma:tau-EY}), $\tau^{\mathrm{EY}}$ can be identified
with the spin relaxation-time under certain assumptions.}
\begin{equation}
\frac{1}{2\tau^{\mathrm{EY}}(k)}=\frac{2\pi}{\hbar}\sum_{\vec k'}
     |\mathcal{H}^{\mathrm{ep}}_{\uparrow\downarrow}(k',k)|^{2}
     \delta(\epsilon_{\uparrow,k}-\epsilon_{\downarrow,k'})
     \label{ma:tau-EY}
\end{equation}
is calculated from~(\ref{ma:H-phon}) with Fermi's Golden
Rule~\cite{pikus-titkov:opt-orient}. As will be discussed in more
detail in~\ref{ma:sub-pol-dynamics}, spin-dependent lifetimes
are usually interpreted as spin
relaxation-times~\cite{pikus-titkov:opt-orient}, i.e., the relaxation
time of the spin \emph{polarization}, but this interpretation is only
valid for a low density of electrons very close to
equilibrium~\cite{DasSarma:lifetime}. Approximately
evaluating~(\ref{ma:tau-EY}) and averaging over $k$ using a Maxwell
distribution for electrons, results in the spin-relaxation
rate~\cite{pikus-titkov:opt-orient}
\begin{equation}
\frac{1}{\tau^{\mathrm{EY}}}
=C\left(  \frac{k_{B}T}{E_{\mathrm{g}}}\right)^{2}\eta^{2}\left(
\frac{1-\eta/2}{1-\eta/3}\right)  \frac{1}{\tau_{\mathrm{p}}} \ .
\end{equation}
Here, $E_{\mathrm{g}}$ is the bandgap energy,
$\eta=\Delta_{\mathrm{SO}}/(E_{g}+\Delta_{\mathrm{SO}})$, \
$\Delta_{\mathrm{SO}}$ the spin-orbit splitting of the holes,
$\tau_{\mathrm{p}}$ the momentum relaxation-time, and $C$ a constant
which equals 2 for the polar interaction with optical
phonons~\cite{song:prb02}. The EY spin relaxation-time is therefore
proportional to the momentum relaxation-time.

\subsection{Dyakonov-Perel Mechanism}

\label{ma:sub-DP}

The Dyakonov-Perel (DP) mechanism~\cite{dyakonov-perel:jetp71} only occurs
in crystals without inversion center because there matrix elements of
$\mathcal{H}_{mn}(\vec{k})$ linear and cubic in $k$ are not forbidden
by symmetry~\cite{pikus-titkov:opt-orient}. Calculating the effective
Hamiltonian for the electronic subsystem taking into account the
spin-orbit interaction in the hole subsystem, one finds in these
crystals that the spin-degeneracy is lifted by the Dresselhaus term,
usually written in the
form~\cite{pikus-titkov:opt-orient,dyakonov-perel:jetp71}
\begin{equation}
\mathcal{H}_{ss'}(\vec{k})=\frac{\hbar}{2}\vec{\Omega}(\vec{k})
\cdot\vec{\sigma}_{ss'}\ .
\label{ma:H-DP}
\end{equation}
In~(\ref{ma:H-DP}), $\vec{\Omega}(\vec{k})$ is a \emph{momentum
dependent} vector (cubic in $k$), and $\vec{\sigma}_{ss'}$ the vector
of Pauli matrices. This effective interaction has the
same effect as an external magnetic field with Larmor frequency
$\vec{\Omega }(\vec{k}),$ which can be calculated by taking into
account the influence of the spin-orbit interaction with the holes
perturbatively. Momentum scattering processes with phonons or
impurities will therefore change the effective magnetic Larmor
frequency~$\vec{\Omega}(\vec{k})$ experienced by an electron by changing its
momentum. An electron therefore experiences a fluctuating magnetic
field, which contributes to the electron spin relaxation. Dyakonov and
Perel first calculated the spin relaxation-time due to this effect in
the collision dominated limit, i.e., when the momentum scattering
time is \emph{shorter }than the time scale of the electron-spin
precession. This situation corresponds to spin relaxation by repeated
small precessions, which is similar to the ``motional narrowing'' in
nuclear magnetic resonance. The DP result is~\cite{pikus-titkov:opt-orient}
\begin{equation}
\frac{1}{\tau^{\mathrm{DP}}}
=\tilde{C}\alpha^{2}
\frac{\left(k_{B}T\right)^{3}}{\hbar^{2}E_{g}}\tau_{\mathrm{p}}
\label{ma:tau-DP}
\end{equation}
where $\tilde{C}$ is a dimensionless factor of order unity depending
on the scattering mechanism and $\alpha$ is a parameter related to the
cubic bandstructure term, given approximately by
\begin{equation}
    \alpha\simeq\frac{4\eta}{\sqrt{3-\eta}}\frac{m_{e}}{m_{0}} \ ,
    \label{ma:alpha}
\end{equation}
and $\tau_{\mathrm{p}}$ the momentum
relaxation-time. In~(\ref{ma:alpha}), $m_{e}$ is the effective
electron mass in the crystal at $k=0$ and $m_{0}$ is the free electron
mass. The spin relaxation-time (\ref{ma:tau-DP}) due to the DP process
is inversely proportional to the momentum relaxation-time.  Therefore,
the resulting spin relaxation-time is \emph{slower} for faster
momentum scattering, leading to the counter-intuitive result that a
\textquotedblleft dirty\textquotedblright\ material with strong
impurity scattering may have a longer spin lifetime.

\subsection{Bir-Aronov-Pikus Mechanism}

\label{ma:sub-BAP}

The Bir-Aronov-Pikus (BAP) process is based on the exchange
interaction between electrons and holes, which can be calculated from
an effective Hamiltonian in 3rd order in the interaction with the
remote bands~\cite{bap:76}. Since we will treat the BAP process in
more depth, and it is a two-particle interaction, we write the
Hamiltonian in 2nd quantization
\begin{equation}
H_{\mathrm{exc}}=\sum_{k_{1},k_{2},q}\sum_{j,s,j',s'}\langle j^{\prime
}s'|\mathcal{V}_\mathrm{exc}(\vec{q})|sj\rangle c_{j',k_{2}+q}^{\dag
}c_{s',k_{1}-q}^{\dagger}c_{j,k_{1}}c_{s,k_{2}} \ .
\label{ma:H-exc}
\end{equation}
Here, $c_{s}$ ($c_{s}^{\dag}$) and $c_{j}$ ($c_{j}^{\dag}$) are
electron and hole destruction (creation) operators [notation as
in~(\ref{ma:js-states})]. The interaction matrix
element~\cite{pikus-bir:71,pikus-titkov:opt-orient}
\begin{equation}
\langle j's'|\mathcal{V}_\mathrm{exc}(\vec q)|sj\rangle=\langle j^{\prime
}s'|\mathcal{V}_{\mathrm{SR}}(\vec q)|sj\rangle+\langle j's^{\prime
}|\mathcal{V}_{\mathrm{LR}}(\vec q)|sj\rangle\label{ma:V-exc}%
\end{equation}
consists of a long-range part 
\begin{equation}
\langle j's'|\mathcal{V}_{LR}(q)|sj\rangle  
=\frac{\hbar^{2}}{m_{0}^{2}}\frac{(\vec{q}\cdot\vec{p}_{j's})
(\vec{q}\cdot\vec{p}_{s'j})}{(\epsilon_{s}-\epsilon_{j'})^{2}}\label{ma:V-LR}
\end{equation}
and a short-range part
\begin{equation}
\langle j's'|\mathcal{V}_{SR}(q)|sj\rangle   
=\frac{3}{2}\Delta_{\mathrm{exc,SR}}\langle j',s'|
\left[  \frac{3}{4}+(\vec{J}\cdot\vec{S})\right]|j,s\rangle \ .
\label{ma:V-SR}
\end{equation}
In~(\ref{ma:V-LR}), the $\vec{p}_{js}$ are the matrix elements of the
momentum operator between the states (\ref{ma:js-states}) defined
in~(\ref{ma:p-matrix-element}). In~(\ref{ma:V-SR}),
$\Delta_{\mathrm{exc,SR}}$ is the excitonic exchange splitting, $\vec
J$ is the hole total angular momentum operator and $\vec S$ is the
electron spin operator. For an explicit matrix representation
of~(\ref{ma:V-LR}) and~(\ref{ma:V-SR}),
see~\cite{maialle:prb96,pikus-bir:71,pikus-titkov:opt-orient}. It
should be noted that~(\ref{ma:H-exc}), (\ref{ma:V-LR}), and
(\ref{ma:V-SR}) are written with the customary definition of
two-particle interaction matrix elements~\cite{reichl:stat-phys}. The
order of the indices shows that~(\ref{ma:H-exc}) describes an
electron-hole exchange interaction process, i.e., an \emph{interband}
scattering process.

Using these contributions, BAP derived a $k$ dependent electron spin
lifetime using Fermi's Golden
Rule~\cite{bap:76,pikus-titkov:opt-orient}
\footnote[2]{The factor
  2 in the following is the same as in~(\ref{ma:tau-EY}).} 
\begin{align}
\frac{1}{2\tau^\mathrm{BAP}_s(k)}=\frac{2\pi}{\hbar}\sum_{\vec{q},\vec{p}}
\sum_{j,j'} &|\langle j's'|\mathcal{V}(\vec{q})|sj\rangle|^{2} 
n_{j,p}(1-n_{j',\vec p -\vec q}) \nonumber \\
&\times \delta(\epsilon_{s,k}+\epsilon_{j,p}
       -\epsilon_{s',\vec k +\vec q}-\epsilon_{j',\vec{p}-\vec q}) 
\label{ma:Gamma-BAP}
\end{align}
where $s'\neq s$ is the flipped spin. As mentioned in
Sec.~\ref{ma:sub-EY}, lifetimes derived from Fermi's Golden
Rule~\cite{pikus-titkov:opt-orient} are usually identified with the
spin relaxation-time, but this identification breaks down for higher
electronic densities or pronounced non-equilibrium
situations~\cite{DasSarma:lifetime}. For a low density
\emph{thermalized} electron distribution, one can thermally
average~(\ref{ma:Gamma-BAP}) with a Maxwell distribution to
obtain~\cite{pikus-titkov:opt-orient,zutic:review}
\begin{equation}
\frac{1}{\tau^\mathrm{BAP}}=\frac{2a_{B}^{3}}{\tau_{0}}
\frac{v_{F}\epsilon_{s,k}}{v_{B}\epsilon_{F}}N_{A}.
\end{equation}
Here, $a_{B}$ and $v_{B}$ are the Bohr radius and velocity of the
exciton, $\epsilon_{F}$ and $v_{F}$ the hole Fermi energy and
velocity, $N_{A}$ the total concentration of holes, and
\begin{equation}
\frac{1}{\tau_{0}} = \frac{3\pi}{64}\frac{\Delta_{\mathrm{exc,SR}}^2}{E_B\hbar}
\end{equation}
where $E_B$ is the excitonic binding energy and
$\Delta_{\mathrm{exc,SR}}$, as defined above, is the excitonic exchange
splitting.

\section{Spin-Polarization \emph{Dynamics }}

\label{ma:sec-polarization-dynamics}

\subsection{Theory of Spin-Relaxation Dynamics}

\label{ma:sub-pol-dynamics}

Electron-spin dynamics in general can be described by the reduced one-particle
density matrix~\cite{sham:mmm99,wu:prb00:kinetics_qw}
\begin{equation}
\rho_{s,s'}=\left(
\begin{array}
[c]{cc}%
\rho_{\uparrow\uparrow} & \rho_{\uparrow\downarrow}\\
\rho_{\downarrow\uparrow} & \rho_{\downarrow\downarrow}%
\end{array}
\right)  \equiv\left(
\begin{array}
[c]{cc}%
n_{\uparrow} & \psi\\
\psi^{\ast} & n_{\downarrow}%
\end{array}
\right)
\label{ma:spin-density-matrix}
\end{equation}
with the real average electronic occupation numbers and the complex
spin-coherence $\psi$ that are driven by a magnetic field. The spin
coherence is only nonzero if the spin-up and spin-down states are no
longer eigenstates of the Hamiltonian, i.e., when the the electrons
are coupled by an external magnetic field or an effective internal
magnetic field as it is the case for the Dyakonov-Perel mechanism. The
dynamics of electrons in a magnetic field therefore follows the
equations of motion for the time-dependent electron distributions
$n_{\uparrow }$, $n_{\downarrow}$ and the spin coherence $\psi$ as
defined
by~(\ref{ma:spin-density-matrix})~\cite{sham:mmm99,hallstein:prb97}. In
semiconductors, both these quantities are subject to carrier-carrier
and carrier-phonon scattering. The microscopic description of the
different interaction mechanisms does not, in general, allow the
introduction of macroscopic $T_{1}$ and $T_{2}$
times~\cite{zutic:review}. \emph{If} these approximate quantities can
be introduced for interacting electrons in semiconductors, $T_{1}$
refers to the relaxation of the macroscopic spin polarization toward
its equilibrium $P=0$ value. For this to happen, the angular momentum
has to be transferred out of the electron system to the holes and
eventually to the lattice. This has to be kept apart from the decay of
the spin coherence which is described by $T_{2}$ at the macroscopic
level. Thus $T_{2}$ is a measure of the dephasing of the electron
spin-coherence driven by magnetic fields and usually called
``homogeneous broadening'' analogous to the case of nuclear magnetic
resonance. If there are additional contributions to the dephasing
related to ``inhomogeneous'' broadening due to, e.g., spatial
fluctuations of the magnetic field, one sometimes describes the
combined effect of the inhomogeneous broadening and the homogeneous
broadening on the decay of the macroscopic spin coherence by
$T_{2}^{\ast}$~\cite{zutic:review}.

In a magnetic field one has two contributions to the time development
of the spin polarization $P$: one is due to the \emph{incoherent}
spin-flip processes related to $T_{1}$ and the other due to the
coherent change of the electronic distribution functions as long as
there is a spin coherence driven by a magnetic field. For the study of
the spin polarization relaxation it can therefore be advantageous to
consider the case without magnetic fields, as will be done here,
because this eliminates the influence of \emph{coherent}
magnetic-field induced spin-flips.  In the following, we will deal
only with the \emph{incoherent} spin-polarization relaxation, i.e.,
the time development of~(\ref{ma:pol}) and its momentum (or
energy) dependent generalization
\begin{equation}
   P_{k}
   =\frac{n_{\uparrow,k}-n_{\downarrow,k}}{n_{\uparrow,k}+n_{\downarrow,k}}.
\label{ma:pol-k}
\end{equation}

A preliminary understanding of the spin relaxation can be achieved
using lifetimes as they result from a Fermi's Golden Rule treatment as
outlined in Sec.~\ref{ma:sub-BAP}. Consider for definiteness the
BAP-process, for which the spin lifetime is described
by~(\ref{ma:Gamma-BAP}). Since Fermi's Golden Rule yields lifetimes,
this is equivalent to a relaxation-time approximation of the
form~\cite{DasSarma:lifetime}
\begin{equation}
\frac{\partial}{\partial t}n_{s}(k)\Bigr|_\mathrm{BAP}
=-\frac{1}{2\tau^\mathrm{BAP}_s(k)}
         [n_{s}(k)-f_s(k)]
\label{ma:ddtn-BAP-out}
\end{equation}
where $f_s(k)$ is an equilibrium (Fermi-Dirac) distribution, to which
the system relaxes. In addition to this relaxation of the electronic
distributions, one has to take into account that the scattered
electrons, which flip their spins during the scattering processes,
have to be accounted for in the number of carriers in the electronic
band with opposite spin. Using this reasoning for a nonzero electronic
spin polarization and an \emph{unpolarized} hole system, i.e.,
$n_{j=\frac{3}{2},k}=n_{j=-\frac{3}{2},k}$ and
$n_{j=\frac{1}{2},k}=n_{j=-\frac{1}{2},k}$, one finds
from~(\ref{ma:Gamma-BAP}) 
\begin{equation}
\tau^{\mathrm{BAP}}_\uparrow(k)=
\tau^{\mathrm{BAP}}_\downarrow (k)\ , 
\end{equation} 
and further obtains
\begin{equation}
   \left.  \frac{\partial}{\partial t}P_{k}\right\vert_\mathrm{BAP}
   =-\frac{P_{k}}{\tau^\mathrm{BAP}_s(k)}\ .
\end{equation}
This result reflects the physical situation that the spin polarization
decays, if the different populations relax towards equilibrium
according to~(\ref{ma:ddtn-BAP-out}), but the scattered electrons show
up in the band with opposite spin with the same momentum after the
scattering process. The information about the energy and momentum
dependence of spin-flip scattering is therefore lost in this
approximation. As will be shown below, the treatment of electron
dynamics as scattering processes without the relaxation-time
approximation leads to a more general description of spin polarization
dynamics due to the exchange interaction. It is interesting to note
that the field of spin relaxation has been dominated by rate-equation
approximations for decades~\cite{opt-orient}, but efforts have been
made recently to go beyond this approximation for the description of
spin-dependent scattering
phenomena~\cite{ivchenko:jetplett02,wu:prb03,wu:prb00:kinetics_qw,roessler:condmat}.

Since the experimental results discussed below are for heavily p-doped
GaAs, the following the discussion will focus on the spin dynamics due
to the electron-hole exchange interaction (\ref{ma:H-exc}), i.e., the
BAP process. To go beyond Fermi's Golden Rule one derives the
dynamical equations for the relevant correlation functions in the
electronic ensemble, which are the distribution functions for
electrons
\begin{equation}
n_{sk}(t)=\langle c_{sk}^{\dag}c_{sk}\rangle,
\end{equation}
defined with time-dependent creation and destruction operators in the
Heisenberg picture. The symbols $\langle\cdots\rangle$ designate a
statistical average over the ensemble of electrons. 
%
%
The dynamics of the carrier distributions $n_{sk}$ is then determined
by the Heisenberg equation of motion
\begin{equation}
\frac{\hbar}{i}\frac{\partial}{\partial t}\langle c_{sk}^{\dag}c_{sk}%
\rangle=\bigl\langle\lbrack H,c_{sk}^{\dag}c_{sk}]\bigr\rangle \ .
\end{equation}
With a two-particle interaction Hamiltonian such as $H_{\mathrm{exc}}$
in~(\ref{ma:H-exc}), one runs into the well-known hierarchy
problem that the equations of motion for the carrier distributions
$n_{sk}$ do not close, i.e., they couple to correlation functions
containing two creation and two destruction operators.  The hierarchy
problem can be approximately solved, e.~g., using Green's
functions~\cite{schaefer-book,binder-koch} or truncation
\cite{mack:pqe99,rossi:rmp02} techniques. Technically, the
random-phase approximation (RPA) will be employed in the
following. One obtains as the dynamical equation for the spin-flip
scattering due to electron-hole exchange interaction a Boltzmann
equation
\begin{equation}
\left. \frac{\partial}{\partial t}n_{sk}\right|_{\mathrm{BAP}}=
-\Gamma_{sk}^{\mathrm{BAP},\mathrm{out}}\, n_{sk}
+\Gamma_{sk}^{\mathrm{BAP},\mathrm{in}}\,(1-n_{sk})
\label{ma:boltz-exchange}%
\end{equation}
with spin and momentum dependent out-scattering 
\begin{align}
\Gamma_{sk}^{\mathrm{BAP},\mathrm{out}}
   = \frac{2\pi}{\hbar}\sum_{\vec{q},\vec{p}}\sum_{j,j'}
   & |\langle j's'|\mathcal{V}_\mathrm{exc}(\vec{q})|sj\rangle|^2 \, 
   n_{j,p}(1-n_{j,\vec p-\vec q})(1-n_{s',\vec k+\vec q}) 
      \nonumber \\
 & \times \delta(\epsilon_{s,k} + \epsilon_{j,p}
-\epsilon_{s',\vec k+\vec q}-\epsilon_{j',\vec p-\vec q})
\label{ma:boltz-out}
\end{align}
and in-scattering rates
\begin{align}
\Gamma_{sk}^{\mathrm{BAP},\mathrm{in}}
= \frac{2\pi}{\hbar}\sum_{\vec{q},\vec{p}}\sum_{j,j'}&
|\langle
j's'|\mathcal{V}_\mathrm{exc}(\vec{q})|sj\rangle|^2 \, 
(1-n_{j,p})n_{j',\vec p-\vec q}n_{s',\vec k+\vec q} 
\nonumber \\
 & \times \delta(\epsilon_{s,k}+\epsilon_{j,p}
-\epsilon_{s',\vec k+\vec q}-\epsilon_{j',\vec p-\vec q}) \ .
\label{ma:boltz-in}%
\end{align}
Here, $s'=-s$ is the flipped spin quantum-number, and the interaction
matrix element is given by~(\ref{ma:V-exc}).  One notices that
the out-scattering rate is identical to the lifetime
obtained by Fermi's Golden Rule (\ref{ma:Gamma-BAP}) if the electronic
occupation of the final state, into which the spin-flipped electron is
scattered, is small, i.e., if $n_{s',\vec k+ \vec q}\ll 1$. Also, a term
mimicking the effect of the in-scattering term in the Boltzmann
equation had to be added to the relaxation-time equation by hand,
since Fermi's Golden rule yields only a lifetime, i.e., a decay time
for individual electrons due to the spin-flipping exchange
interaction. 

Formally, a similar Boltzmann equation describes the time development
of the hole distribution functions under the action of the
electron-hole exchange scattering. As will be shown later
experimentally and theoretically, the spin-flip
scattering~(\ref{ma:boltz-exchange}) due to the electron-hole exchange
interaction takes place on a timescale of several ten picoseconds in
moderately to strongly p-doped GaAs. We are here exclusively
interested in p-doped GaAs where a large density of holes and a very
small density of electrons is present. Thus the hole angular momenta
equilibrate due the rapid momentum scattering and spin-orbit
interaction much faster than the electronic spins. We will therefore
treat the holes only as a bath in the following calculations.

In addition to the electron-hole exchange interaction, carriers
interact also via the \emph{direct} Coulomb interaction. This leads
to strongly carrier-density dependent scattering times that
reach several hundred femtoseconds for high carrier densities. We
therefore have to include the effect of the fast direct electron-hole
Coulomb scattering, which is spin conserving, in addition to the
spin-flip electron-hole exchange scattering since the electronic
distribution functions in~(\ref{ma:boltz-exchange}) will evolve
under the action of the direct scattering quickly. This is described
by the well-known Boltzmann equation for the direct Coulomb scattering
in random-phase approximation~\cite{haug-koch}
\begin{align}
\left.  \frac{\partial}{\partial t}n_{sk}\right\vert _{\mathrm{Coul}} 
=-\frac{2\pi}{\hbar}\sum_{k'qj}|W_{q}|^{2}&\bigl[n_{sk}
(1-n_{sk+q})n_{jk'+q}(1-n_{jk'})\nonumber\\
&\mbox{}  -(1-n_{sk})n_{sk+q}(1-n_{jk'+q})n_{jk'}\bigr]\nonumber \\
&\mbox{}\times\delta(\epsilon_{jk+q}+\epsilon_{sk}
-\epsilon_{jk'}-\epsilon_{sk+q}) \ .
\label{ma:boltz-direct}
\end{align}
Here, $W_{q}=V_{q}/\varepsilon_{q}$ is the screened Coulomb
interaction in momentum space defined in terms of the bare Coulomb
interaction
\begin{equation}
  V_q = \frac{e^2}{\varepsilon_0 \varepsilon_{\mathrm{bg}}q^2}
\end{equation}
with the background dielectric function
$\varepsilon_{\mathrm{bg}}$. The carrier contribution to the
dimensionless dielectric function is calculated in static
approximation $\varepsilon_{q}=1/(1 + q^2/\kappa^2)$ where $\kappa$ is
the screening wavevector~\cite{haug-koch}.

Equations~(\ref{ma:boltz-exchange}) and~(\ref{ma:boltz-direct}) define
our model that will be used to calculate the dynamics of electronic
distribution functions and thus the spin relaxation. We will neglect
electron-electron Coulomb scattering because the electronic densities
will be kept so low that this process is much slower than the
electron-hole Coulomb scattering. We will also use fixed Fermi-Dirac
distributions for the high-density hole distributions because the fast
hole-hole Coulomb scattering together with the spin-orbit interaction
keeps the hole system in thermal equilibrium and prevents the
occurence of hole spin polarization on timescales important for the
electronic spin dynamics.

\begin{figure}[hbt]
\begin{center}
\resizebox{0.6\textwidth}{!}{\rotatebox{270}{\includegraphics{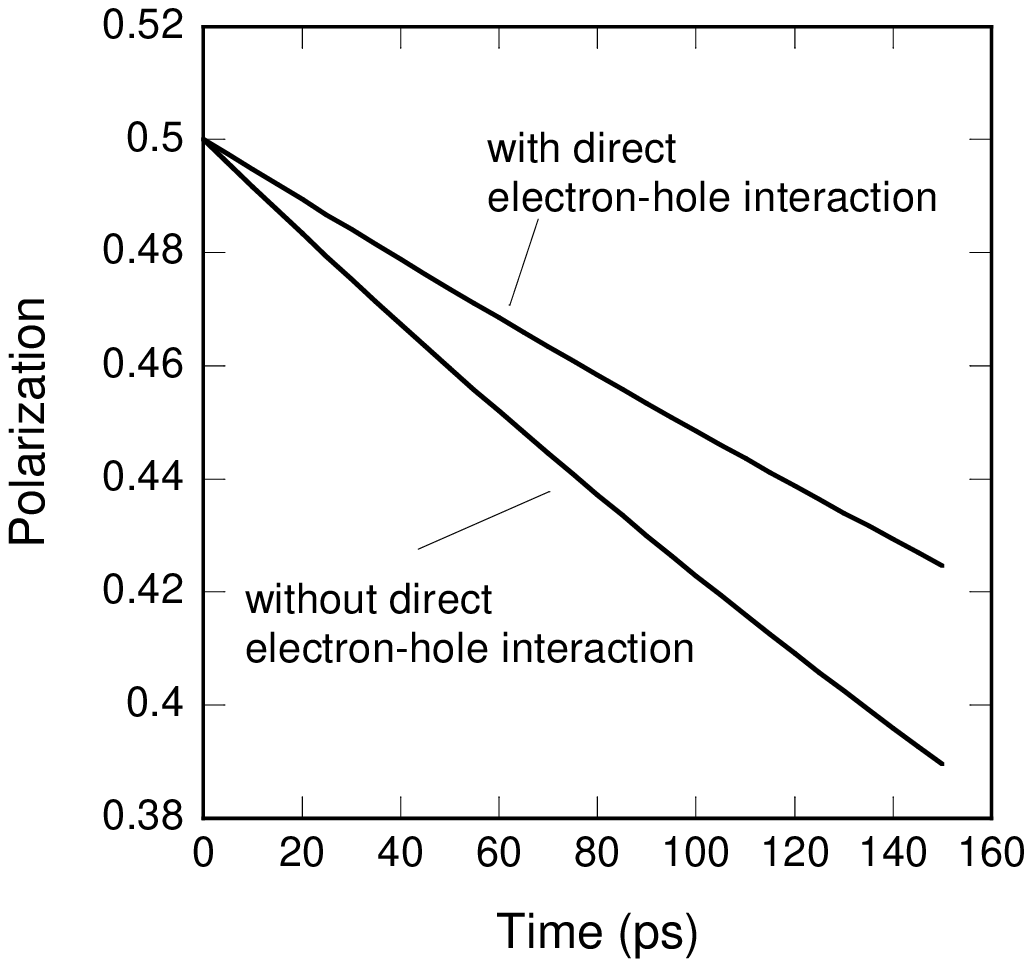}}}
\caption{Computed relaxation of electron spin polarization $P(t)$ for
electron-hole exchange interaction with direct Coulomb interaction
(top curve) and exchange interaction without Coulomb interaction
(bottom curve). At $t=0$, carrier temperatures, densities and spin
polarizations are 300K, $10^{18}\,\mathrm{cm}^{-3}$ and 0\% for holes, and
600\,K, $10^{15}\,\mathrm{cm}^{-3}$ and 50\% for electrons,
respectively.}%
\label{ma:fig-dir-exc}%
\end{center}
\end{figure}

\subsection{Numerical Results}

\label{ma:sub-numerical-results}In this section results of the numerical solution
of~(\ref{ma:boltz-exchange}) and~(\ref{ma:boltz-direct}) are
presented.  The numerical solution is accomplished by transforming the
sums into integrals and then using a 4th order Runge-Kutta algorithm
to calculate the time evolution of the electronic distribution
functions. The initial conditions in the following are always taken to
be quasi-equilibrium distributions for electrons and holes. After
optical excitation of spin polarized carriers in p-doped GaAs the
electrons will quickly equilibrate thermally without losing their spin
coherence on this timescale, whereas the holes will lose their spin
polarization due to spin-orbit interaction and Coulomb scattering so
that one can assume quasi-equilibrium conditions with polarized
electrons and unpolarized holes.

\begin{figure}[t]
\begin{center}
\resizebox{0.6\textwidth}{!}{\rotatebox{270}{\includegraphics{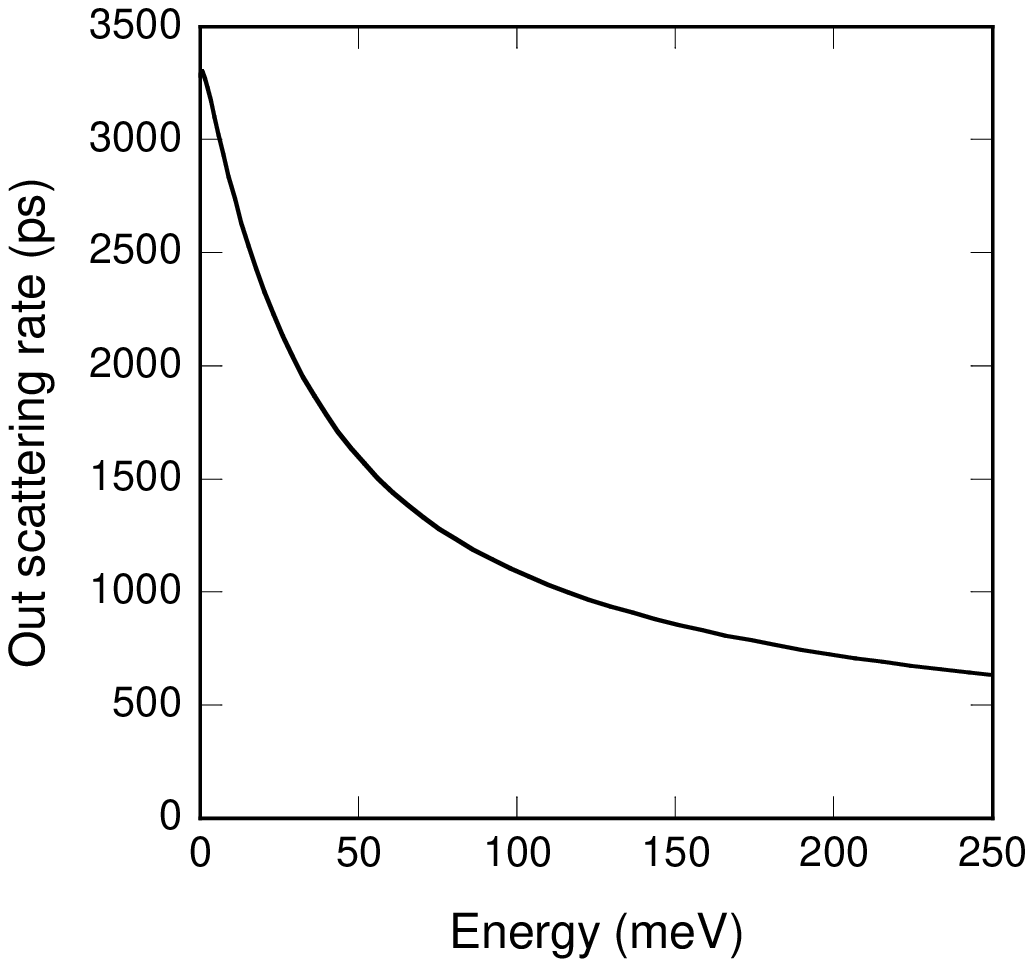}}}
\caption{Calculated out-scattering rate~(\protect\ref{ma:boltz-out}) vs.\
electronic energy for the same parameters as in
Fig.~\protect\ref{ma:fig-dir-exc}.}
\label{ma:fig-outscatt}
\end{center}
\end{figure}

As a first result of the model we will discuss the interplay between
direct Coulomb scattering and exchange
scattering. Figure~\ref{ma:fig-dir-exc} shows the result of a numerical
experiment, in which the case with Coulomb scattering is compared to
the one where Coulomb scattering is ``switched off.''For the case of a
polarized low electron density at 300 K in the presence of an high
density of unpolarized holes at 600 K, one finds that the direct
Coulomb interaction has a significant influence on the dynamics of the
electron spin polarization over several tens of picoseconds, even
though the characteristic timescale of direct Coulomb scattering is on
the order of a few hundred femtoseconds. Thus the direct scattering
influences the dynamics of the electron spin polarization, even though
it does not give rise to spin flips. The reason for this behavior is
that the exchange scattering alone creates non-equilibrium electron
distributions over time, whereas the direct scattering continually
drives the electronic distributions into equilibrium with the holes.
The resulting competition leads to the faster electron
spin-polarization decay if both scattering mechanisms are included. It
should also be noted that, even though the time-development of the
total polarization is plotted here, the time development of the
energy-resolved spin polarizations are not much different from the
total polarization. In the range of electron and hole temperatures
around room temperature, the energy-resolved polarization and the
total polarization can be fitted with an exponential decay law, and
the time constants are practically equal for all those quantities. In
the case of Fig.~\ref{ma:fig-dir-exc} we find a relaxation time of 950
ps for the electronic polarization.

Figure~\ref{ma:fig-outscatt} shows the energy-dependent out-scattering
rate~(\ref{ma:boltz-out}) for electrons due to the exchange
interaction with unpolarized holes, with all the parameters being the
same as before. It should be noted that the out-scattering rate is
identical to the Fermi's Golden Rule result in the low
electron-density limit. This quantity was calculated earlier for
$T=0$\,K~\cite{maialle:prb96} as a measure of the spin relaxation
time, which was predicted to be strongly energy dependent on the basis
of this calculation.
\begin{figure}[t]
\begin{center}
\resizebox{0.6\textwidth}{!}{\rotatebox{270}{\includegraphics{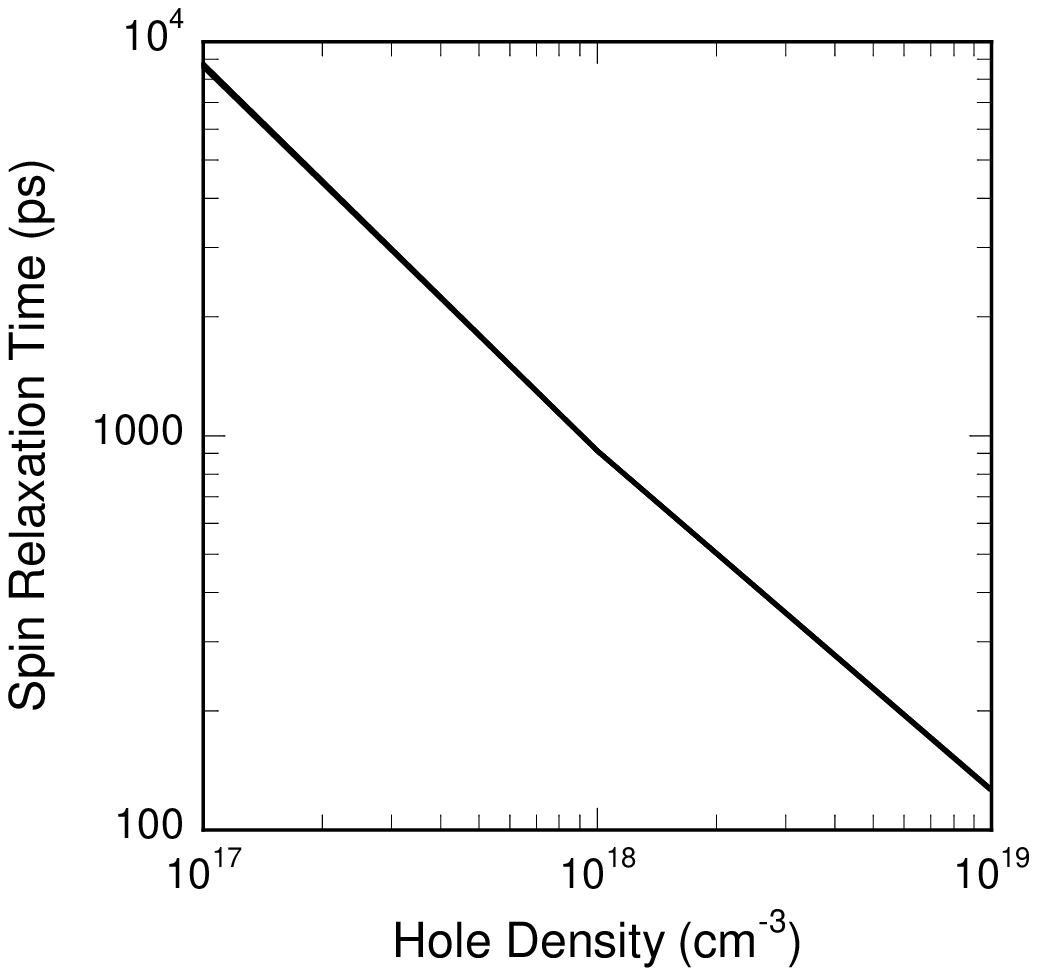}}}
\caption{Calculated spin relaxation-time vs.\ density of holes for
hole temperature of 300\ K, electron density $10^{14}\,\mathrm{cm}^{-3}$
and temperature 600 K. The initial polarization is again $P(t=0)$ =
50\%.}
\label{ma:fig-trelax-N}%
\end{center}
\end{figure}
However, as mentioned in connection with Fig.~\ref{ma:fig-dir-exc}, no
energy-dependence of the spin-polarization relaxation is found in our
numerical results including the dynamics due to both exchange and
direct scattering processes.

Figure~\ref{ma:fig-trelax-N} shows the computed spin
relaxation-time for different hole concentrations. For the range of
densities depicted in Fig.~\ref{ma:fig-trelax-N}, the dynamics of
electron spins was calculated over 180 ps, and an exponential fit was
made to the resulting polarization dynamics. The result is a
meaningful measure of the effectiveness of the BAP process for the
polarization relaxation that can be compared to experiments.  For a
high doping concentration of $N_{A}=10^{19}\,\mathrm{cm}^{-3}$ we obtain a spin
relaxation time of 110\,ps, which is in very good agreement with recent
measurements of the bulk spin relaxation in identical samples using
time-resolved Faraday rotation
techniques~\cite{beschoten:gan-spincoherence:prb01}. In principle, one
can therefore use the calculated density dependence of the
spin relaxation-time to obtain an estimate of the hole density and
thus the doping concentration in a GaAs/metal interface (Schottky
barrier), where the band bending near the surface leads to a depletion
of holes in the vicinity of the surface. Before we turn to the
discussion of the experimental setup and the experimental results, we
briefly review the concept of a Schottky contact and the band bending
at semiconductor-metal surfaces.

\subsection{Spin Decay in a Schottky-Barrier}

\label{ma:sub-schottky-theory}

Most metal-semiconductor interfaces act as a diode because the
electric current passing through the interface depends exponentially
on the forward bias since a depletion layer in the semiconductor
builds up~\cite{balluffi,moench:book}. These contacts are known as
``Schottky contacts,'' as opposed to ohmic contacts which have a lower
resistance. The concept of a depletion layer and the accompanying band
bending is most easily described for a semiconductor vacuum surface
before moving on to the actual Schottky
contact~\cite{balluffi}.

\begin{figure}[t]
\begin{center}
\resizebox{0.5\textwidth}{!}{\includegraphics{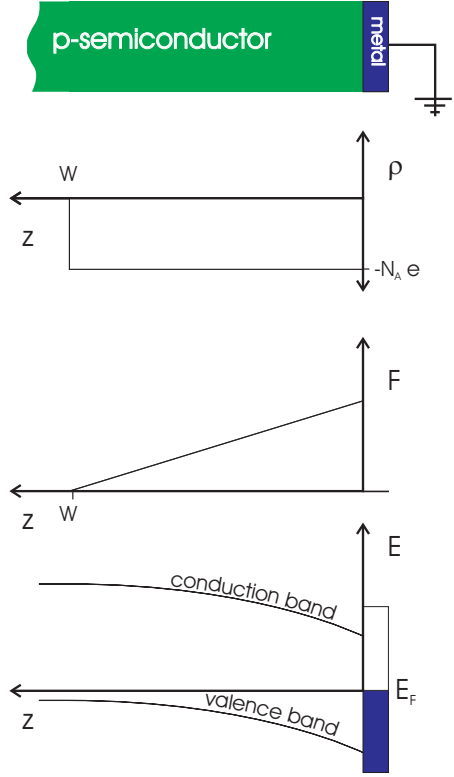}}
\caption{Schottky contact between p-doped semiconductor and metal. The charge
distribution $\varrho$ is approximated as a step function, leading to
the electric field distribution and band lineup shown below.}%
\label{ma:fig-schottky}%
\end{center}
\end{figure}

Consider a uniformly p-doped semi-infinite slab of semiconductor
material, where the majority of carriers are holes, and the doping
concentration is $N_{A}$. The introduction of a surface breaks the
crystal symmetry and leads to additional carrier states located near
the surface, whose energies lie in the semiconductor bandgap and which
are occupied predominantly by holes in a p-doped semiconductor. The
holes occupying the surface states lead to a positive surface charge
$\sigma$ and must have come from acceptors in a region of width $w$
beneath the surface, the so-called depletion layer, where there is now
a negative space charge $\varrho=-eN_{A}$ per unit volume, see
Fig.~\ref{ma:fig-schottky}. (Here, $e>0$ denotes the magnitude of the
electronic charge.) Charge neutrality requires
\begin{equation}
\sigma + \varrho w =0 .
\end{equation}
The positive space charge of the depletion layer now causes an electrostatic
potential, which is responsible for the characteristic band bending. This, in
turn, reduces the surface charge and compensates it almost completely. To see
this, one calculates the electrostatic potential $\Phi(z)$ away from the
surface induced by the space charge of the depletion layer according to
Poisson's law
\begin{equation}
\frac{\D^{2}}{\D z^{2}}\Phi(z)
=-\frac{\varrho}{\varepsilon_{b}\varepsilon_{0}} \ ,
\end{equation}
where $\varepsilon_{b}$ and $\varepsilon_{0}$ are the permittivity of the
semiconductor and the vacuum, respectively. In solving the equation for
$\Phi(z)$ we assume that at the end of the depletion layer ($z=w$) we can set
the electrostatic potential $\Phi$ and the field
$\mathcal{F}=-\D\Phi/\D x$ to zero. 
We find a simple linear
dependence for the electric field%
\begin{equation}
\mathcal{F}(z)=-\frac{\varrho}{\varepsilon_{b}\varepsilon_{0}}(w-z)
\end{equation}
and for the potential accordingly
\begin{equation}
\Phi(z)=-\frac{\varrho}{2\varepsilon_{b}\varepsilon_{0}}(w-z)^{2}.
\end{equation}
Thus the additional potential energy for an electron with charge
$q=-e$ in the the region $z<w$ is $-e\Phi(z)$, leading to a downward
band-bending.

With this background discussion one can turn to the
metal-semiconductor Schottky contact~\cite{balluffi}, and consider
again the case of a p-type semiconductor. It is assumed that the metal
has a chemical potential, which is denoted by the Fermi energy
$E_{F}$, and the semiconductor has a chemical potential $\mu$. Before
bringing the two materials into contact the respective chemical
potentials are different, with the semiconductor chemical potential
lying in the bandgap close to the valence band because of the
p-doping, and therefore lower than the metal chemical
potential. Bringing the two materials together sets up an electric
field due to the different chemical potentials which drives electrons
into the semiconductor, thereby creating a negative space charge in the
semiconductor and a positive surface charge, just as in the example
discussed above. When the materials are joined, these two charge
distributions balance each other.  Also, the common chemical potential
is pinned because the surface (or interface) states are only partially
filled. The negative space charge in the semiconductor leads to the
band bending described above, as shown in Fig.~\ref{ma:fig-schottky}.

When carriers are created in the semiconductor away from the surface
by optical excitation near the bandgap, this band configuration is the
reason that diffusion of optically excited electrons from the bulk
towards the surface yields electrons with a high kinetic energy, i.e.,
``hot'' electrons, at the surface~\cite{pierce:photoemission}.

\section{Experiments}

\subsection{Time and Spin Resolved 2PPE}

Carrier spin-relaxation measurements in zero applied field have been
reported in~\cite{crooker:prb97} by means of time-resolved Faraday
rotation experiments. In order to isolate the spin decay occurring
only in the band-bending region, a surface sensitive and energy
resolving technique is needed. To explore the spin- and charge
dynamics in this region we introduce a novel real-time method. We use
a pump-probe approach referred to as time-resolved
2-photon-photoemission (TR-2PPE) with the additional option to measure
the spin-polarization of the emitted electrons~\cite{ma:prl97}. By
varying the time delay between the ultrashort pump and probe laser
pulses, the spin-dependent population decay of the intermediate
(unoccupied) states can be determined. The high surface sensitivity as
well as the energy selectivity of the photoemission technique is
appropriate to investigate the spin dynamics in the Schottky
barrier. From this perspective, TR-2PPE is complementary to Faraday
rotation. In TR-2PPE one simply determines the spin polarization~$P$
defined in~(\ref{ma:pol-k}) of the transient carrier concentration in a
certain energy interval. The energy-resolved spin polarization~$P$ is
independent of the carrier populations at that energy. A reduced
carrier concentration in the investigated energy interval will only
lead to an increased statistical error on the measured spin
polarization.

\begin{figure}[t]
\begin{center}
\resizebox{0.7\textwidth}{!}{\includegraphics{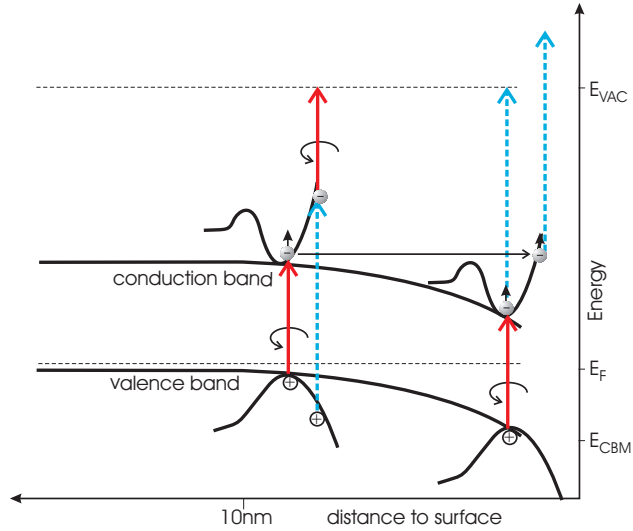}}
\caption{Bichromatic 2PPE-process in a surface region with band
bending. By making use of the energy resolution of the photoemisssion
technique, we can probe the hot electron spin dynamics at different
energy levels. Parts of the band structure at different distances from
the surface are sketched. The solid arrows mark the transitions
induced by the absorption of 1.55\,eV pump photons, and the dashed
arrows the transitions induced by the 3.1\,eV probe photons.}%
\label{ma:fig:2ppe-band-bending}%
\end{center}
\end{figure}

In order to study the spin decay in the band bent region by TR-2PPE,
we first excite spin-polarized electrons with a circularly polarized
fs-laser pulse at $\hbar\omega=1.55$\,eV, slightly larger then the
bandgap. Some of the spin-polarized hot electrons will move towards
the band-bending area, where the carriers become hot (see
Fig.~\ref{ma:fig:2ppe-band-bending}) and will undergo different
elastic and inelastic relaxation processes. The spin dependent
depletion of the excited-state population in the surface region can be
determined by a suitably delayed second laser pulse which photoemits
the electrons. The second (probe) laser pulse must be at higher photon
energy ($\hbar\omega = 3.1$\,eV) in order to overcome the vacuum
energy.

\subsection{Experimental Setup}

\label{ma:sub-experimental-setup}

\begin{figure}[t]
\begin{center}
\resizebox{0.8\textwidth}{!}{\includegraphics{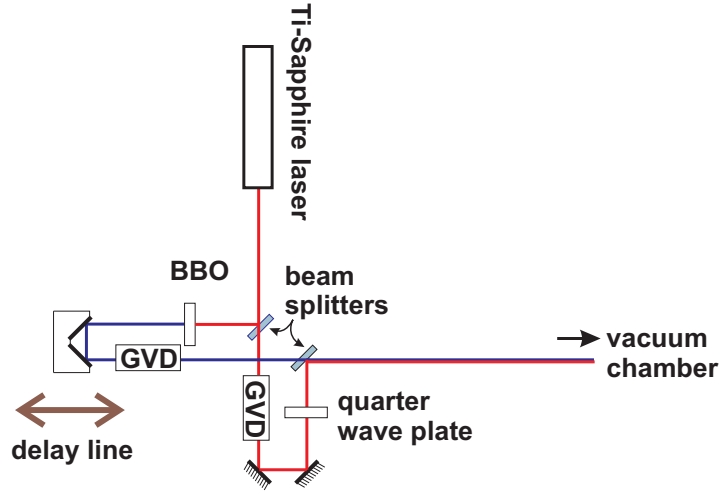}}
\caption{Schematic set-up of the Mach-Zehnder-interferometer. The
pulses are split up and can be delayed with respect to each other. In
addition, we can modify each arm of the interferometer to tailor the
beam properties to our specific needs by introducing e.g. frequency
doubling elements or quarter wave plates. Each arm then needs
individual dispersion compensation.}%
\label{ma:fig-exp-setup}
\end{center}
\end{figure}

The laser system used is by now a standard tool for measurements of
the type described here. It consists of a mode locked Titanium-doped
Sapphire (Ti:Al$_{2}$O$_{3}$) laser, pumped by a diode laser at about
8\,W. This setup generates transform-limited and sech$^{2}$ temporally
shaped pulses of up to 15 nJ/pulse with a duration of less than 45 fs
at a repetition rate of 82\,MHz.  The wavelengths of the pulses can be
tuned from 830 to 770\,nm, thereby varying the photon-energy from
1.49\,eV to 1.61\,eV. For the time resolved experiments the pulse
train is split by a beam splitter (see
Fig.~\ref{ma:fig-exp-setup}). By varying the optical path with a
variable delay line, we can shift the two pulse trains by a certain
length corresponding to a delay in time.

The pulse of one path is frequency doubled by a thin
$\beta$-barium-borate-(BBO-)crystal to yield photon energies in the
range of 3.0 to 3.2\,eV (in the following designated by
\emph{blue}). The pulse of the other path (fundamental, in the
following designated by \emph{red}), is transformed from linearly
polarized to circularly polarized by introducing a quarter-wave
plate. Adjusting the quarter wave plate, we can change between
left-handed and right-handed light.  This setup is used for the
optical orientation experiments. The frequency-doubled pulses remain
linear. After delaying, the pulses are re-united by a second beam
splitter and then focused by a lens onto the sample inside the
UHV-chamber. The polarized laser beam is incident perpendicular on the
sample surface, and the electrons are detected at an angle of
45$^{\circ}$ with respect to the surface normal.
In order to account for the pulse broadening introduced into the
system by dispersive elements, a group velocity dispersion
compensation is necessary for both pulse trains. This is accomplished
with a prism pair traversed twice by the laser pulse resulting in
negative dispersion, which cancels out the usually positive dispersion
due to lenses and the vacuum viewport.

The laser-beams are adjusted to reach excitation densities as low as
$1\times10^{16}\,\mathrm{cm}^{-3}$. The excitation densities are of the
same order of magnitude for the fundamental laser beam and for the
frequency doubled pulses, although the intensities are approximately
100 times smaller for the 3.1\,eV pulses. This is due to the fact that
the penetration depth for 3.1\,eV photons is 50 times smaller than for
the 1.55\,eV photons. Considering the laser intensities used and the
absorption lengths $\lambda$ at a wavelength of 800\,nm
($\lambda_{r}$= 730\,nm) and 400\,nm ($\lambda_{b}$ = 14\,nm) in GaAs,
we can roughly estimate the average excitation density from the laser
power. Assuming a Gaussian profile for our leaser beam, we determine
the focus spot size of our lens to 100\,$\umu$m in diameter. Taking the
maximum laser powers available, we can reach excitation densities on
the order of $5\times10^{18}$ electron-hole pairs cm$^{-3}$. By
reducing the pulse-power (reducing the laser output and/or lowering
the frequency doubling efficiency), more than three orders of
magnitude in excitation density are available.

The time-averaged photocurrent at a fixed kinetic energy is measured
as a function of the delay between the two beams (two pulse
correlation technique).  The nonlinearity of the two-photo process
leads to an increase in the 2PPE yield when the pulses are spatially
and temporarily superimposed. As long as the two laser pulses
overlap in time, it is obvious that an electron can be emitted by
absorbing just one photon from each pulse. However, if the pulses are
separated in time an excited electron from the first pulse
can absorb a photon from the second pulse only as long as the
inelastic lifetime of the intermediate state exceeds the delay.

The samples are mounted in a UHV chamber (base pressure
$5\times10^{-11}$ \,mbar) equipped with a cylindrical sector electron
energy analyzer. To investigate the electron dynamics separately for
both spin directions, a spin analyzer
(SPLEED-detector~\cite{ma:prl97,kirschner-contrib}) is mounted on top
of the electron-energy analyzer. This makes the measurement of one
in-plane component of the spin-polarization vector possible. A bias
voltage of $-14$\,V is applied to the sample to eliminate the effects of
any stray electric and magnetic fields.

As sample we use a p-doped (100) oriented GaAs crystal with dopant
concentration (zinc) $N_{A}=1\times10^{19}$cm$^{-3}$. GaAs surfaces
are quite reactive when exposed to air and the surface of a sample
stored outside a vacuum chamber is oxidized. The samples are cleaned
by etching in sulfuric acid and rinsing, followed by heating
to approximately 500$^{\circ}$C in the UHV chamber. Prior to the
measurements the sample is treated with a small amount of cesium in
order to obtain a well-defined Fermi-level pinning and a lowered
work function~\cite{moench:book}.

\subsection{Experimental Results for GaAs (100)}

\label{ma:sub-exp-GaAs}

\begin{figure}[t]
\begin{center}
\resizebox{0.7\textwidth}{!}{\includegraphics{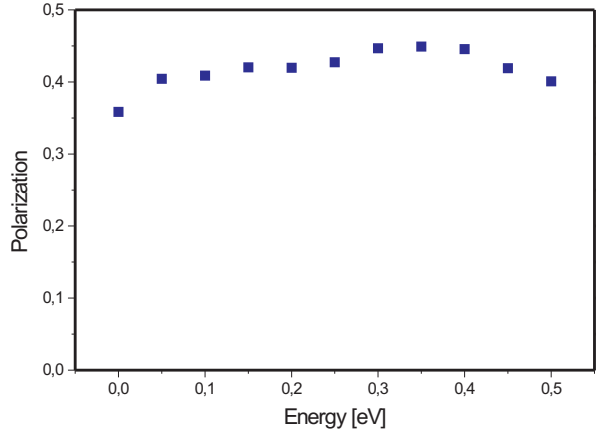}}
\caption{Spin polarization of photoexcited electrons ($\hbar\omega$ =
1.55~eV), injected into the (100) surface of a GaAs-crystal
with band bending. The zero of the energy scale is defined as the conduction
band minimum at the surface ($E_{\mathrm{CBM}}$) and represents the
energy of the intermediate state after absorption of the first
photon.}%
\label{ma:fig-pol-vs-E}
\end{center}
\end{figure}

By making use of the energy resolution of the photoemisssion technique
one can probe the spin polarization of the hot electrons at different
energy levels in the band-bending region. Figure~\ref{ma:fig-pol-vs-E}
shows the spin polarization versus energy at zero delay time when
there is temporal overlap of the pump and probe pulses.  The reference
energy is the conduction band minimum (CBM) at the surface
($E_{\mathrm{CBM}}$). We observe nearly uniform spin polarization
within the band-bending energy range. Taking into account that the
electrons close to the conduction band minimum have undergone many
elastic and inelastic scattering processes, the result indicates that
the spin relaxation in the band-bending region must occur on a much
longer time scale than the energy relaxation-time. This has been
questioned previously by several authors considering hot
electrons~\cite{rasing:apb99,opt-orient,pierce:photoemission}.

In the following time dependent measurements we keep the energy
of the detected photoelectrons constant and vary the temporal delay
between the exciting pulse (which yields the spin polarization) and
the probing pulse. As discussed in
Sec.~\ref{ma:sub-experimental-setup}, we use the bichromatic 2PPE
method, where pump (\emph{red}) and probe (\emph{blue}) laser have
different photon energies.

\begin{figure}[b]
\begin{center}
\resizebox{0.7\textwidth}{!}{\includegraphics*{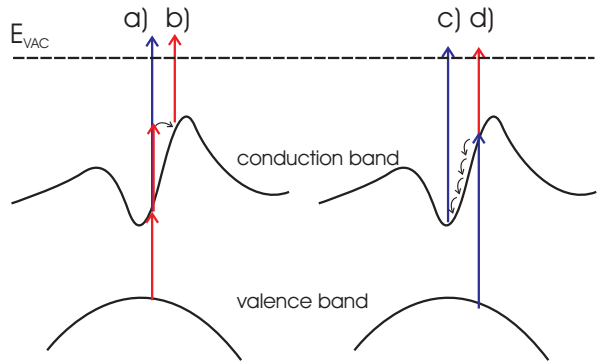}}
\caption{Multiphoton processes: a) \emph{red-blue} transition
yielding spin polarization if the \emph{red} pulse is circularly
polarized. b) \emph{triple red} transition yielding also spin
polarization. c) \emph{blue-blue} transition involving inelastic
scattering events. d) \emph{blue-red} transition. Both c) and d)
yield zero spin polarization because the \emph{blue} pulse is
linearly polarized. These events occur in overlapping energetic
regions, when the experiment is carried out in the band-bending
surface area.}%
\label{ma:fig-multiphoton-processes}%
\end{center}
\end{figure}

In a bichromatic experiment electrons can be excited in two different
ways (see Fig.~\ref{ma:fig-multiphoton-processes}). With photons of
1.55\,eV average energy and circular polarization (\emph{red}) we
create a spin aligned population with a spin-polarization of 50\% in
theory or approximately 40\% at room temperature. These electrons are
excited with a small excess energy near the $\Gamma$ point (see
Figs.~\ref{ma:fig-multiphoton-processes}a
and~\ref{ma:fig-multiphoton-processes}b). The second kind of electrons
are excited by linearly polarized photons of 3.1\,eV (\emph{blue})
leading to a non-polarized and highly excited population in the
conduction band (see Figs.~\ref{ma:fig-multiphoton-processes}c and
~\ref{ma:fig-multiphoton-processes}d). In a second step this
population may gain additional energy by other photons. Let us
consider electrons absorbing enough energy to reach the vacuum level,
which can be properly adjusted by a treatment of the surface with
alkali atoms such as Cs.  It is obvious that the bichromatic
transitions (\emph{red-blue} and \emph{blue-red} transitions) are
delay dependent since they involve photons of both pulses. The
monochromatic contributions (\emph{blue-blue} or \emph{triple red})
are delay independent since they occur independently of the other
laser pulse. Hence, monochromatic transitions will dominate the signal
at long delay times ($\varDelta t\rightarrow\pm\infty$), when the pulses
are well separated, whereas the bichromatic events will dominate at
delay times within the energetic relaxation time of the electron.

\begin{figure}[tb]
\begin{center}
\resizebox{0.8\textwidth}{!}{\includegraphics{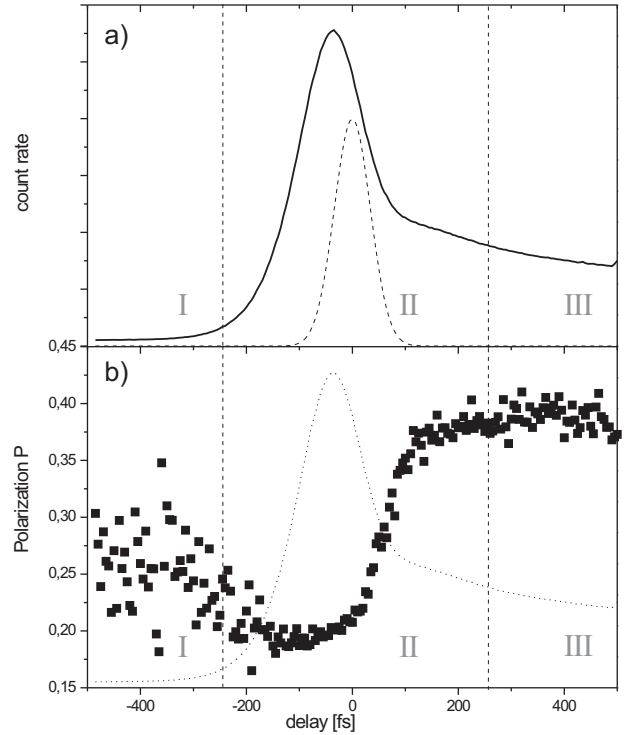}}
\caption{(a) 2PPE photo-current as a function of the pump-probe delay
time.  The dotted curve shows the cross-correlation of the
laser-pulses (photon-fluence). (b) Spin dynamics around short delays,
the dotted curve is a replica from (a). We can distinguish 3
regimes. From left to right: I.  \emph{blue-red} regime: The low count
rate is mainly given by background electrons from 3-photon-absorption
(see Fig.~\ref{ma:fig-multiphoton-processes}).  II. Overlapping regime:
Spin drops drastically due to highly excited unpolarized electrons,
which decay very fast. III. \emph{Red-blue} regime: constant spin
polarization on the ps-timescale.}%
\label{ma:fig-exp-countrate}%
\end{center}
\end{figure}

The first event in these multiphoton processes is a direct interband
transition and, therefore, not only the intermediate, but also the
initial and final state will be energetically distinguishable between
the two processes (see Fig.~\ref{ma:fig-multiphoton-processes}). This
results in a different kinetic energy of the \emph{blue-red} and the
\emph{red-blue} transition photoelectrons and they can be separated by
means of an energy analyser. In the band bending region, however, we
cannot energetically discriminate between the \emph{red} excited
populations and the \emph{blue} excited electrons completely. The
energy of the initial state (relative to the vacuum energy) varies as
a function of the distance to the surface, resulting in partly
overlapping energy ranges of \emph{red-blue} and \emph{blue-red}
events, see Fig.~\ref{ma:fig:2ppe-band-bending}.
Next to the intermediate state close to the conduction band minimum
(probed due to a \emph{red-blue} process) one also probes a higher
lying state due to the \emph{blue-red} process. Therefore, working in
the bichromatic mode one has to consider, that the signal is given by
the dynamics of both probed intermediate state. In a semiconductor,
however, the dynamics of both probed intermediate states do not
interfere with each other. Figure~\ref{ma:fig-exp-countrate}(a) shows the
2PPE photocurrent as a function of the pump-probe delay time. Positive
(negative) delay time corresponds to \emph{red-blue} (\emph{blue-red})
transitions. The dotted curve shows the cross correlation of the laser
pulses (photon fluence) given by the 2PPE yield vs.~delay time on a
transition metal surface with very short relatation times. The result
shows that the part of the measured signal due to \emph{blue-red}
transitions, which transit via the higher excited intermediate states
(see Fig.~\ref{ma:fig-multiphoton-processes}), shows a much faster
population decay compared to the \emph{red-blue} part, which probes
the dynamics close to the conduction band minimum. Hence, due to the
large difference in the population decay between \emph{red-blue} and
\emph{blue-red} transition, we can distinguish very easily between
these two (in general interfering) 2PPE processes. The simultaneously
measured spin-polarization of the photoelectrons is illustrated in
Fig.~\ref{ma:fig-exp-countrate}(b). We can divide our
polarization data curves into three different regimes:

\begin{description}[Regime III:]
\item[Regime I:] Excitation with linearly polarized 3.1\,eV and
emission with circular 1.55\,eV-photons at a negative delay of 250 fs
or more. The comparatively weak signal is dominated by background
electrons such as spin-polarized 3-photon-processes of 1.55\,eV
photons (see Fig.~\ref{ma:fig-multiphoton-processes}b) and unpolarized
electrons of 2-photon-processes of 3.1\,eV photons (see
Fig.~\ref{ma:fig-multiphoton-processes}c) which lost energy in a scattering
event.  This regime is characterized by a low count rate and therefore
high noise in spin signal.
\item[Regime II:] overlapping regime: the pulses overlap in time and,
hence, bichromatic transitions dominate. The left shift of the 2PPE
cross-correlation curve compared to time zero indicates that
unpolarized \emph{blue-red} transitions dominate over polarized
\emph{red-blue} transitions, resulting in a strongly reduced spin
polarization of the detected photoelectrons. In this region we find a
strong signal and very accurate spin determination.
\item[Regime III:] \emph{red-blue} transition at a delay $>$
200\,fs. The region with the highest spin polarization, since the
signal is dominated by optically induced spin polarized electrons
close to the conduction band edge. These intermediate states are
long-lived leading to a high count rate over a wide delay
range. Observing the spin polarization with increasing delay time
allows the determination of the spin relaxation-time $T_{1}$ as long
as the energetic relaxation is not much faster than the spin
relaxation.
\end{description}

\begin{figure}[tb]
\begin{center}
\resizebox{0.6\textwidth}{!}{\includegraphics{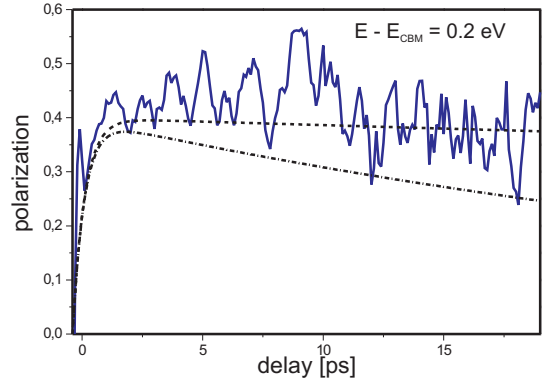}}
\caption{Spin polarization vs.~delay time for hot electrons at 0.2\,eV
above conduction band minimum at the surface. The dashed curve
corresponds to a spin decay on the order of 300\,ps. The dash-dotted
curve represents a spin decay time of 60\,ps.}%
\label{ma:fig-exp-pol-200meV}%
\end{center}
\end{figure}

In the following we will discuss the spin decay processes at the
conduction band minimum and 200\,meV
above. Figure~\ref{ma:fig-exp-pol-200meV} shows the spin polarization
versus positive delay time (\emph{red-blue} transitions) for hot
electrons at 200\,meV above the conduction band minimum. The excess
energy is a multiple of the longitudinal optical phonon energy
$\hbar\omega _{p}=36$\,meV, but it is lower than the split-off
valence-band energy $\Delta_{\mathrm{SO}}=340$\,meV. The data were
corrected for the decaying electron population under the assumption
that the background contributions from the two pulses
(\emph{blue-blue} and \emph{triple red}) remain constant for all
delays. Since the transient population of excited electrons decays
quite fast on a time scale of a few 100\,fs, the noise level increases
considerably over delay time. The polarization shows no spin
relaxation on our time scale of about 20\,ps in length. The
dash-dotted curve is a fit with $T_{1}$ = 60\,ps, which represents the
GaAs bulk value expected from our theoretical analysis and
experimentally verified for the same sample by means of time-resolved
Faraday rotation at $T=300$\,K. The fact that the spin polarization
remains constant over the investigated delay time indicates that all
additional quasi-elastic scattering processes in the band-bending
region, e.g.. at steps, defects, and impurities do not cause spin flips.
Compared to the bulk, the spin relaxation of hot electrons in the
interface is slower.

\begin{figure}[bt]
\begin{center}
\resizebox{0.6\textwidth}{!}{\includegraphics{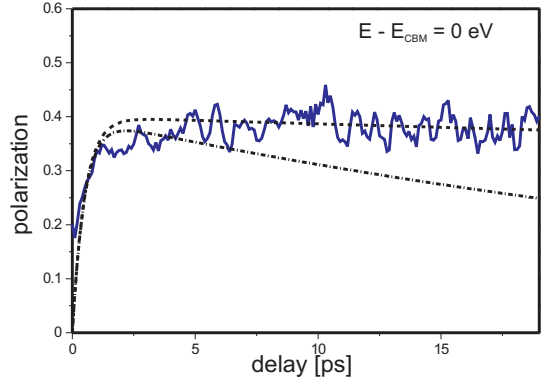}}
\caption{Spin polarization versus delay-time at conduction band
minimum. As a guide to the eye, the dashed curve corresponding to a
spin decay on the order of 300\,ps. The dash-dotted curve represents
the bulk value of $T_{1}$ = 60\,ps, as obtained by a time-resolved
Faraday rotation experiment.}
\label{ma:fig-exp-pol-bandminimum}
\end{center}
\end{figure}

In Fig.~\ref{ma:fig-exp-pol-bandminimum} the spin-polarization versus
positive delay time is shown for electrons at the conduction band
minimum of the surface. No direct excitation is possible using photon
energies at 1.55\,eV (bandgap energy at room temperature:
$E_{\mathrm{g}}=1.42$eV). Time-resolved measurements show that the
population increases at zero delay and reaches its maximum at a delay
of 100 fs. Hence, the electrons in this energy interval have been
generated either in the band-bending region at the surface or 
in the bulk and have diffused towards the surface. The carriers
have undergone many inelastic scattering processes, none of them
includes a spin-flip process as clearly shown in
Fig.~\ref{ma:fig-exp-pol-bandminimum} for longer time delays.

Again, the spin relaxation rate in the interface is decreased compared
to the bulk value (dotted line) as determinded by the time-resolved
Faraday rotation experiment. Our experimental data show a lower limit
for $T_{1}$ of more than 300\,ps. It should be noted that for p-doped
semiconductors a much faster population decay of the carriers compared
to the spin decay makes the exact determination of $T_{1}$ difficult
by means of a real time experiment.

\subsection{Comparison between Experimental and Theoretical Results}

For the p-doping concentration considered here, the BAP process,
i.e., the electron-hole exchange interaction, is considered to be the
dominating contribution to spin relaxation~\cite{song:prb02}. The
reason for the experimental result is shown by our calculations: To
obtain reliable spin relaxation-times one has to compute the time
development of the distribution functions for spin-up and spin-down
electrons including the relevant scattering mechanisms and accounting
for in and out-scattering events, thus going beyond existing
evaluations of the spin relaxation at the level of Fermi's Golden
Rule~\cite{maialle:prb96,song:prb02}. In the case of strongly p-doped
GaAs these scattering mechanisms are electron-hole exchange
scattering, which is comparatively slow but can flip the electron
spins, and electron-hole direct Coulomb scattering, which is much
faster and cannot flip spins. Because the exchange scattering alone
does not lead to a thermalization between electrons and holes, the
direct scattering process, which does lead to this thermalization, is
needed to provide physically sensible results if the electronic
distributions are tracked over several tens of picoseconds. From the
time development of the electronic distribution functions one then
calculates the spin relaxation and determines the relaxation time by
an exponential fit. This procedure yields energy-dependent spin
polarizations whose relaxation times have only a very weak energy
dependence, in contrast to the rates leading to the spin relaxation.

Since the experimental results presented above were obtained using
2PPE, where the spin polarization is measured for electrons from the
surface or interface region of the semiconductor, we can assess the
spin relaxation in the semiconductor/metal interface
region. The electrons that are detected in the interface region have
been optically excited in the bulk of the semiconductor by the
\emph{red }laser pulse and have then reached the interface region with
strong band bending. It has been argued that the electrons are
subject to efficient scattering process when they reach the band
bending region where the become ``hot'' and thus that their
spin-polarization should relax more quickly than in the
bulk.\cite{pierce:photoemission} Our experimental results show the
opposite effect: The spin relaxation-time is significantly
longer than the corresponding value for electrons in the bulk. The
explanation for this behavior is that the electrons in the interface
region are effectively separated from a high hole density because they
become effectively localized in the potential well formed by the
downward slope of the electron band and the surface. In the surface
region, on the other hand, the concentration of holes is strongly
reduced because of the depletion of holes close to the interface in a
Schottky barrier. This explanation is supported by our calculations:
We see from Fig.~\ref{ma:fig-trelax-N} that the relaxation time is
indeed very sensitive to the hole concentration, and that the
relaxation time thus is a measure of the hole concentration in the
interface region.  Using the experimental value of 300 ps relaxation
time, we obtain the rough extimate of a 40\% decrease in hole
concentration at the surface from Fig.~\ref{ma:fig-trelax-N}.

For the comparison between theory and experiment presented above, we
have taken the theoretical bulk spin relaxation-time as a
measure of the spin relaxation at the interface and we have allowed
the hole concentration to vary over several orders of magnitude. For
lower doping concentrations, it is expected that a contribution from
the DP process will also come into play and some details of the
scattering mechanisms can also be affected by the localization of the
electrons close to the surface, which essentially restricts them to
move in a two-dimensional layer. There is as yet no straightforward
explanation for the reduced DP process at the interface. One may
assume a preferential orientation of the conduction electron momentum
in the (100) direction of the GaAs interface caused by the electric
field in the band bending region. Taking into account that there is no
spin splitting of the conduction band along the (100) axis of
GaAs~\cite{opt-orient}, this alignment effect can reduce the spin
relaxation due to the precession along the internal field. A
quantitative theoretical analysis of the influence of the DP process,
as well as an inclusion of the reduced dimensionality of the electron
gas at the interface are left for future investigations

We thank B. Beschoten for helpful discussions and sharing his
experimental results before publication. We are grateful to M.~Bauer,
M.~Fleischhauer, W.~H\"ubner, and S. W.~Koch for helpful
discussions. Financial support from the DFG and a CPU grant from the
Forschungszentrum J\"ulich are gratefully acknowledged.

\bibliographystyle{unsrt}
\bibliography{dephasing2,spin3}


\end{document}